# Systematic comparison of deep belief network training using quantum annealing vs. classical techniques

Joshua Job[1] and Steve Adachi[1, *]

[1]*Lockheed Martin Advanced Technology Center, Sunnyvale, CA 94089*[†]

(Dated: July 16, 2020)

In this work we revisit and expand on a 2015 study that used a D-Wave quantum annealer as a sampling engine to assist in the training of a Deep Neural Network. The original 2015 results were reproduced using more recent D-Wave hardware. We systematically compare this quantum-assisted training method to a wider range of classical techniques, including: Contrastive Divergence with a different choice of optimizer; Contrastive Divergence with an increased number of steps (CD-k); and Simulated Annealing (SA). We find that quantum-assisted training still outperforms the CD with Gibbs sampling-based techniques; however, SA is able to match the performance of quantum-assisted training trivially using a quench-like schedule with a single sweep at high temperature followed by one at the target temperature.

Deep Learning technology has made tremendous progress within the last decade, from detecting human faces and cats in images, [1] to surpassing human ability on certain tasks such as playing Go.[2] However, these advances have placed ever-increasing demands on computing hardware, which has evolved from CPUs to GPUs to special purpose processors, especially for the computationally intensive task of training deep neural networks. In the future, this evolution may lead to hybrid computing architectures including quantum processors.

Such a hybrid quantum-classical training method for deep neural networks was previously studied by one of the authors [3] as well as by others.[4] This method uses a quantum annealer, such as the devices made by D-Wave Systems, [5] as a sampling engine during the training process. It was found that on a small-scale test case (coarse-grained MNIST images), this method achieved a higher accuracy than the conventional training approach based on Contrastive Divergence (CD).[6] However, the original study was purely empirical and did not shed any light on whether the advantage of this "quantum-assisted training" method was due to quantum effects, or whether it could be reproduced by more sophisticated classical techniques.

In the current work, we confirm that the 2015 results can be reproduced using more recent D-Wave hardware. We also systematically compare the quantum-assisted training method to a wider range of classical techniques, including: Contrastive Divergence with a different choice of optimizer; Contrastive Divergence with an increased number of steps (CD-k); and Simulated Annealing (SA). [7] We found that quantum-assisted training still outperforms CD even using a different optimizer (e.g. Adam [8]), or with the number of steps increased (e.g. CD-1000). However, we found that SA is able to match the performance of quantum-assisted training. In fact, SA performed well even when the number of sweeps was reduced from 1000 down to just 2. Thus, the apparent advantage seen previously for quantum-assisted training cannot be ascribed to quantum effects, as there is an efficient classical algorithm that can accomplish the same results.

---

* steven.h.adachi@lmco.com
† joshua.job@lmco.com

Since both CD and SA can be viewed as thermal sampling methods, it is an open question why SA with just two sweeps performs so much better than CD-1000 on this problem. This is an interesting question which could merit further study. Nevertheless, further investigation of CD vs. SA would not alter the primary finding of this paper, namely that there does not appear to be an advantage using present day (stoquastic, transverse-field Ising model) quantum annealing devices to train *classical* RBMs in this way. While this is a discouraging result for this approach to quantum machine learning, we note our work here has no bearing on applications of such quantum annealers when employed as quantum Boltzmann machines nor the potential utility of annealers with non-stoquastic couplings [9–11]. We hope that the work presented here will be useful to other investigators as an example of best practices for benchmarking quantum computing devices, particularly for machine learning. Before we can draw conclusions whether a "quantum advantage" exists or not in some scenario, it is generally not sufficient to compare with a single classical technique, and we must always ask ourselves whether there are better classical techniques available, as illustrated here.

## QUANTUM ANNEALING FOR DEEP LEARNING

This paper focuses on an approach for training a Deep Neural Network (DNN) that was first studied by Hinton et al. [12] This approach is based on a Deep Belief Network (DBN), a multilayer model where each layer is a Restricted Boltzmann Machine (RBM) [13]. The DBN is first trained generatively using Contrastive Divergence (CD), [6] then the weights are fine-tuned using backpropagation [14].

CD is an approximation that was introduced in order to avoid exact computation of the gradient of the log-likelihood, which involves expectation values over $2^N$ configurations, where N is the number of nodes in one RBM layer. However, there is an inherent error in not following the gradient, and it can take many iterations of generative training using CD to get good results using this method. For example, training on the MNIST data set [15] took about a week on a 3GHz Xeon processor to achieve results comparable to feedforward neural networks [12]. The approximation error can be reduced by taking k contrastive divergence steps per training iteration (CD-k), but this takes more computation time, and in practice the most common variant is 1-step Contrastive Divergence (CD-1).

Since the RBM formulation is based on a quadratic energy functional over binary variables, it can easily be mapped to the problem Hamiltonian of a quantum annealer. Then the intractable terms in the gradient can be estimated using samples from a quantum annealer. This hybrid quantum-classical approach, which we call **quantum-assisted training of neural networks**, was studied by one of the authors [3] as well as by Benedetti et al.[4] A similar approach was also studied in the context of gate-based quantum computing by Wiebe et al.[16]

The empirical results in [3] using the D-Wave hardware indicated that this quantum-assisted training method achieved greater accuracy in fewer training iterations than purely classical Contrastive Divergence based training, on a coarse-grained version of the MNIST data set. An example of these results is shown in Figure 1.

In the figure, the horizontal axis shows the number of pre-training iterations, while the vertical axis shows the accuracy on the test data set. The red curves are the results for classical training using CD-1, while the blue curves are the results for the quantum-assisted training. The figure on the left is for 100 post-training iterations, while the figure on the right is for 800 post-training iterations. (For full details of the experiment, see [3].)

A number of questions were left unanswered by this earlier study. In particular, is the apparent advantage of quantum-assisted training due to quantum effects, or can it be explained classically? Also, are there other classical techniques besides contrastive divergence that could match the performance of quantum-assisted training? These are the questions we set out to answer in the current effort.

The scope of the project included the following:

- The original 2015 quantum-assisted training code was updated to run on the more recent D-Wave 2X quantum annealer at the University of Southern California (USC), and the quantum-assisted training experiment on the coarse-grained MNIST data set was repeated to confirm whether the same behavior could be observed on the newer hardware and associated software.

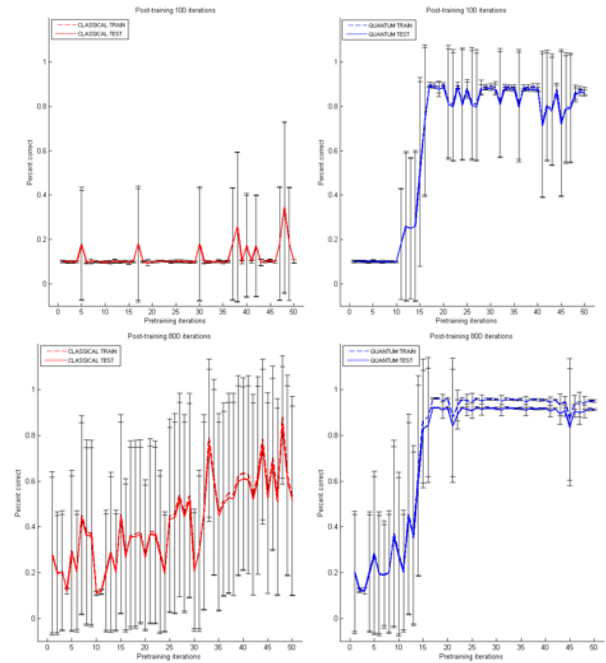

FIG. 1. Prior results from Adachi & Henderson (2015) showing significant advantage in training speed and accuracy for QA-based pretraining over CD.

- Comparisons were made against a wider variety of classical techniques, including the following:
  - Different choice of optimizer (Adam)
  - Increased number of Contrastive Divergence steps (CD-k instead of CD-1, up to k=1000)
  - Better Gibbs samplers, e.g. Simulated Annealing (SA) and Parallel Tempering (PT) [17]

## METHODS, ASSUMPTIONS, AND PROCEDURES

The quantum-assisted training method, as well as all of the classical techniques we use for benchmarking, all use a common Deep Neural Network model and data set. Details specific to the quantum-assisted training method have been documented previously in [3], but are summarized again below. Finally, we describe the overall strategy for benchmarking and determination of quantum effects, and the classical techniques we use for that purpose.

### Deep Neural Network Model

Here we briefly describe the basic model used in our tests and the fundamentals of the training algorithm so as to make explicit where the various algorithms we test will fit in.

A Deep Belief Network (DBN) is a multilayer model where each layer is a Restricted Boltzmann Machine (RBM) consisting of "visible" and "hidden" nodes connected in a bipartite graph, as illustrated in Figure 2.

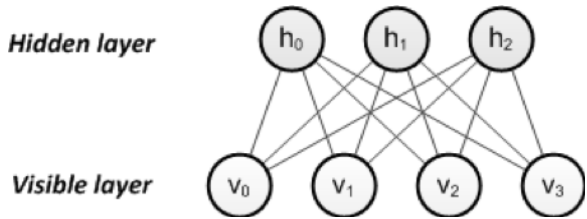

FIG. 2. Restricted Boltzmann Machine

The RBM joint probability distribution is defined by a Gibbs distribution

$$P(v,h) = \frac{1}{Z} \exp\left(-E(v,h)\right) \quad (1)$$

with an energy functional

$$E(v,h) = -\sum_{i=1}^{n} b_i v_i - \sum_{j=1}^{m} c_j h_j - \sum_{i=1}^{n}\sum_{j=1}^{m} W_{ij} v_i h_j \quad (2)$$

for $v_i, h_i \in \{0,1\}$ where $n$ is the number of visible nodes and $m$ is the number of hidden nodes. The normalization constant

$$Z = \sum_{\{v_k\}} \sum_{\{h_l\}} \exp\left(\sum_k b_k v_l + \sum_l c_l h_l + \sum_{kl} W_{kl} v_k h_l\right) \quad (3)$$

is known in physics as the partition function.

Because of the bipartite graph structure, the forward and reverse conditional probability distributions for an RBM are both simple sigmoid functions:

$$P(h_j = 1 \mid v) = \text{sigm}\left(c_j + \sum_i W_{ij} v_i\right) \quad (4)$$

$$P(v_i = 1 \mid h) = \text{sigm}\left(b_i + \sum_j W_{ij} h_j\right) \quad (5)$$

In this study, we use a DBN consisting of a 32 node input layer, two hidden layers of 32 nodes each, and a 10 node output layer. This can be viewed as 3 stacked RBMs, with respective sizes 32x32, 32x32, and 32x10.

Generative training of the DBN is done by greedy layer-wise training of each RBM. The lowest RBM layer is trained using the real data for the visible units. The remaining layers are trained by sampling from the hidden layer of the preceding RBM and using the samples as the data for the next higher RBM.

The goal of RBM training is to determine values of the parameters {W,b,c} that maximize the likelihood (1) conditioned on the visible data $\{v_i\}$. It is customary to work with $\log P$ and to compute the gradients $\partial_{W_{ij}} \log P$, $\partial_{b_i} \log P$, and $\partial_{c_j} \log P$, which can be expressed as:

$$\frac{\partial \log P}{\partial w_{ij}} = \langle v_i h_j \rangle_{\text{data}} - \langle v_i h_j \rangle_{\text{model}} \quad (6)$$

$$\frac{\partial \log P}{\partial b_i} = \langle v_i \rangle_{\text{data}} - \langle v_i \rangle_{\text{model}} \quad (7)$$

$$\frac{\partial \log P}{\partial c_j} = \langle h_j \rangle_{\text{data}} - \langle h_j \rangle_{\text{model}} \quad (8)$$

The "positive phase", ie the terms $\langle v_i h_j \rangle_{\text{data}}, \langle v_i \rangle_{\text{data}}, \langle h_i \rangle_{\text{data}}$, are all easily computed as they are expectations given the data. For each element in the dataset, the visible units are fixed, and we can use $P(h|v)$ from above to exactly compute the expectations. A simple average over the training data then yields the full expectation over the data. The "negative phase", the expectations over the model, are intractable to compute exactly. In general for an n x m RBM (n visible units and m hidden units) the time to compute the model expectations scales as $2^{\min(n,m)}$ as one must take a sum over all states in one of the layers (naively this would be $2^{n+m}$ but one can exactly integrate out the larger of the layers at the expense of inducing interactions within the remaining layer).

At each training iteration, the weights and biases {W,b,c} are updated by applying an optimizer function using the previous values of {W,b,c} and the gradients above as inputs. In the previous work [3], a momentum-based optimizer was used, but other optimizers such as Adam [8] could also be used.

Regardless of the choice of optimizer, the difficult computational task is to compute the model expectations $\langle v_i h_j \rangle_{\text{model}}$, $\langle v_i \rangle_{\text{model}}$, and $\langle h_i \rangle_{\text{model}}$. Since doing so exactly is intractable, all work with RBMs resorts to some sort of heuristic. In the quantum-assisted training, these expectations are estimated using sampling from the D-Wave quantum annealer. Below, we will consider various classical techniques which approximate these expectations using some variant of thermal Markov Chain Monte Carlo (MCMC) sampling.

**Data Set**

The quantum and classical techniques studied here all use the same data set previously studied in [3], which is a coarse-grained version of the MNIST data set of handwritten digits [15]. The MNIST images were reduced in size due to the limited size of RBM that can be mapped onto the D-Wave architecture. Images were reduced in size as described in [3]. As in the original data set, there are 60,000 training images and 10,000 test images. Each image has a truth label from 0 to 9.

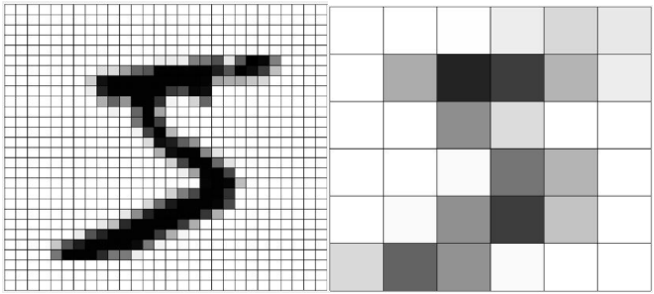

FIG. 3. Example of Original (28x28) and Coarse-Grained (6x6) MNIST Images

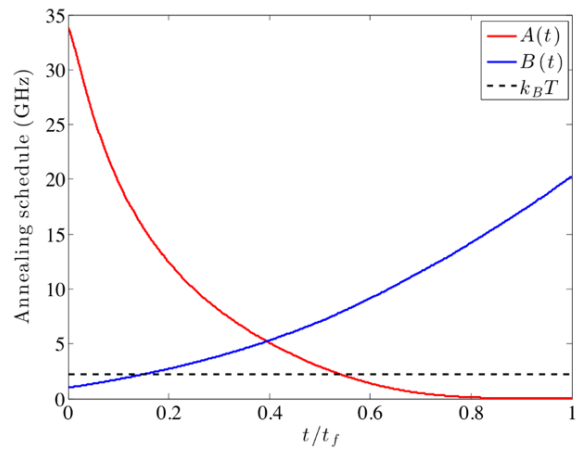

FIG. 4. Annealing Schedule with Envelope Functions A(s), B(s)

## Quantum-Assisted Training

*Quantum Annealing.*

Quantum annealing [18] is a method which is inspired by simulated annealing and similar classical methods. Rather than initializing a system at a high temperature and slowly dropping the temperature to reach a ground state or a thermal state, one instead initializes the system in a quantum state corresponding to an equal superposition of all classical bit-strings and slowly modifies the Hamiltonian of the system so that it eventually matches the classical energy function of your problem.

Mathematically, the Hamiltonian as a function of time t is given by

$$H(s) = A(s)\sum_i \sigma_i^x + B(s)\sum_i \left(h_i + \sum_j J_{ij}\sigma_j^z\right)\sigma_i^z \quad (9)$$

A(s) is a function that starts large and falls to zero while B starts small and rises, see Figure 4 where $s = t/t_f$.

*Mapping From RBM to the D-Wave Quantum Annealer*

If one takes the computational basis to be that of $\sigma^z$, one can immediately see a connection between the energy function E(v,h) of the RBM and the $\sigma^z$ component of the QA Hamiltonian. Indeed, they are identical in form, and the QUBO function of the RBM can be easily mapped into the Ising form of QA and vice versa.

In practice, things are not quite so straightforward. The actual QA device from D-Wave Systems [5] has a limited architecture, requiring us to map a chain of qubits (ie nodes) in the physical annealer into a single node of the RBM, a process known as minor embedding. Moreover, the DW device operates at a physical temperature which is not adjustable, and has limited precision in programming the weights and biases. In addition, from the above curve for A(s) and B(s) one can predict that the system will effectively stop evolving long before the end of the anneal schedule. When the energy scale for the classical Hamiltonian becomes much larger than both device temperature and the quantum component of the Hamiltonian, the system essentially stops evolving altogether. This means that the state of the system we measure is not a Gibbs state for the final, classical Hamiltonian we are interested in for RBM training, but at best a quantum Gibbs state corresponding roughly to the time when quantum perturbations became small. This was explored in [19].

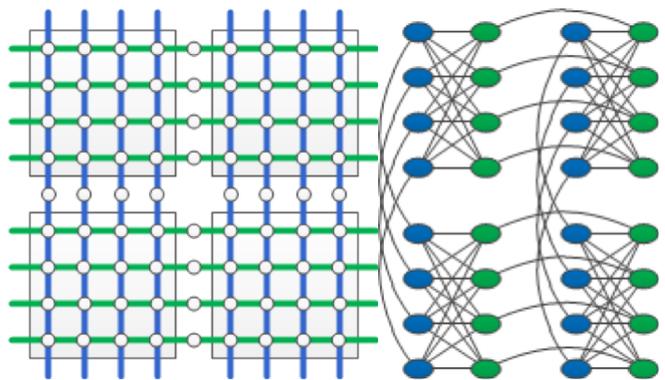

FIG. 5. Minor Embedding Using Qubit Chains for Visible (Blue) and Hidden (Green) Nodes

Finally it has been observed that when attempting to use the D-Wave device to sample from a classical RBM Gibbs distribution, there is an additional parameter, the effec-tive temperature (not the physical temperature) which is problem dependent and must be estimated. We use the temperature estimation technique described in [3]; another technique has been described in [4].

All this is meant to say that the mapping from RBM to QA is not as direct as it may appear at first glance, and any attempt to demonstrate a true quantum advantage or origin for the results of the original study will have to proceed through the steps from logical RBM to physically embedded quantum annealing system. The research plan based on this is outlined in the next section.

*Updates to Quantum-Assisted Training Since the 2015 Paper*

To reproduce the 2015 results [3] on Lockheed Martin's current D-Wave system, a number of updates were needed to the prior quantum-assisted training code. These updates included:

- Updates due to hardware changes. The prior system was a D-Wave 2 (504 working qubits) whereas the current system is a DW2X (1098 working qubits). So, changes to the code were made to reflect the increased number of qubits as well as the locations of "bad" qubits.
- Updates due to software changes. The code was updated to use the D-Wave SAPI 3.0 API.
- Effective temperature was re-estimated due to differences in hardware operating temperature and noise characteristics between the two machines.
- Updates for bias measurement and correction. D-Wave instituted an automated hourly flux drift compensation process (colloquially known as "shimming"), which caused the statistics of samples to change over time. D-Wave subsequently provided a flag to turn off this process, so the code was updated to set this flag to false, and to use an alternative bias measurement and correction process as described in [20].

In addition, some minor bugs in the original code were found and fixed. Most notably, a bug was fixed in the original code that generated the 32-pixel coarse-grained MNIST data set. Experiments reported here all used the corrected data set. Fortunately, the differences in the results between the earlier incorrect data set and the corrected data set, appear to be insignificant.

**Benchmarking vs. Classical Techniques and Determination of Quantum Effects**

The basic outline for the research program conducted here, as shown in Table I, is to test, in sequence, the potential origins of the observed advantage for the QA-based pretraining observed in the original study from least to most interesting (here approximately synonymous with quantum-ness). The advantage may have been a simple fluke or accident, if this is the case simply trying the classical training again would likely reveal the problem. It may be that QA was doing a better job of approximating the true Gibbs state than CD-1. If this is the case, it may be that we merely need to run contrastive divergence for longer, or use a better Gibbs sampler like SA or PT. It may be some interplay between the minor embedded problem and the dataset, which we would discover if we could only recover the observed performance for QA using classical samplers on the embedded problems. And finally there may be some real quantum effect going on, either at the single-site or whole-system level, which we can discover if we were forced to go to algorithms such as spin-vector Monte Carlo (essentially a system with no entanglement but coherent qubits) or path integral Monte Carlo (a truly quantum system thermalizing in the instantaneous energy eigenbasis, which is the expected behavior of a quantum annealer such as D-Wave in the weak coupling limit without error suppression/correction). In outline, this is the same procedure recommended in [21].

Testing began at the top of the table, exhausting one raft of explanations before proceeding to the next. Ultimately, after a number of rounds of iteration, we were not required to proceed past row 2 of Table 1. After running CD with different optimizers and observing no statistically significant difference, we ruled out the option that it was a fluke. Details of this will be provided below. To test performance from Gibbs samplers, CD was run for 1,2,10,100, and 1000 steps with batch size 100, and SA was run for a number of sweeps running from 2, 10, 100, 1000, and 10000 with the same number of samples per step as D-Wave, namely 400 samples each. It was also tested for two sweeps with 100 samples per step, both for fixed and random ordering for variable updates.

*Contrastive Divergence (CD)*

Estimating the model expectations uses various heuristics. Hinton [6] introduced the contrastive divergence method, which samples $h|v_{\text{data}}$ for each element in the dataset and then repeatedly samples v|h and h|v using their exact marginal distributions for k total rounds of joint sampling. This procedure, for parameter k, is called CD-k. In [3], QA was compared against CD-1 as a model expectation heuristic.

CD-k does not exactly approximate the gradient of the system except when $k \to \infty$ but in general "works" even for small k (typically CD-1 is used).

| Origin of QA advantage | How we can test it | How we might potentially achieve it with classical systems |
|---|---|---|
| Fluke (ie isn't robust) | Try again, test other data sets, etc. | N/A |
| QA yields better approximations of true Gibbs distribution | Try better Gibbs samplers than CD-1; ie CD-1000, SA, PT, etc. | Highly optimized, GPU Gibbs samplers |
| Embedding logical problem introduces implicit prior on the logical distribution | Test advanced Gibbs samplers on embedded problems | Highly optimized, GPU Gibbs samplers |
| Genuine quantum distribution yields better implicit prior | SVMC Path Integral Quantum Monte Carlo | Large cluster of GPUs, computationally expensive |

TABLE I. Potential origins of observed QA advantage

*Simulated Annealing and Related Classical Thermal Samplers*

A more advanced technique for estimating model expectations would be to use a general purpose algorithm for sampling from Ising models, of which an RBM is one. Simulated annealing (SA), which initializes a Markov chain randomly and then performs Metropolis updates, ie changing the state {v,h} to {v',h'} with a probability of $\max\left(1, \exp\left(-\beta\left(E\left(v', h'\right) - E\left(v, h\right)\right)\right)\right)$. As the inverse temperature $\beta = 1/T$ is slowly raised, ie the temperature T is slowly lowered from an initial effectively infinite temperature (meaning all updates are accepted) to a very low temperature (meaning essentially only updates which decrease energy are accepted) one can reach the ground state of an Ising model, given sufficiently many steps. This can also be used to thermalize an Ising model at a temperature of 1, which is what is needed for RBM training. In essence, one can envision SA as a sequence of MCMC runs used to approximate the probability density of a sequence of Ising models. At high temperature, the densities are very smooth and easy to approximate, and if one adjusts temperature down slightly the new density will be somewhat less smooth but close enough to the prior density that a Markov chain initialized with the previous run will quickly converge at the new lower temperature.

In essence, rather than, as in CD, starting from data and exploring only a small region around the data to approximate a gradient, SA starts from a very simple but related model and slowly changes its model to match the one of interest.

*Other Methods*

More advanced techniques, such as Parallel Tempering (PT), may also be used, and often converge more quickly. PT essentially runs multiple Markov chains at the same time at different temperatures, occasionally exchanging the state of the chains across temperatures via a Metropolis update. It is typically the best performing thermal sampling algorithm available for classical models.

## RESULTS AND DISCUSSION

**Quantum-Assisted Training**

The results of quantum-assisted training using the updated code are shown in Figure 6. These are very similar to the corresponding quantum-assisted training (blue curves) in Figure 1 for 100 and 800 post-training iterations respectively.

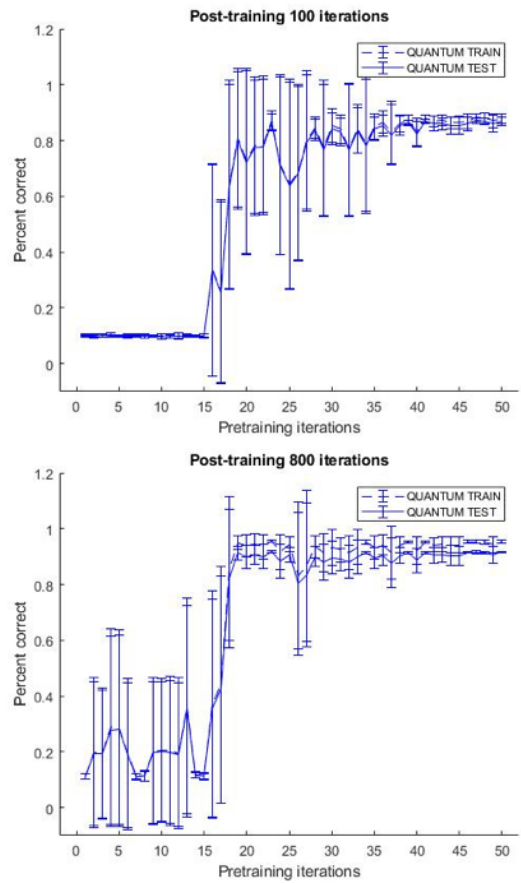

FIG. 6. Previous Quantum-Assisted training results reproduced using updated code, showing similar behavior as previously.

Thus it appears that the behavior observed previously is relatively robust to the changes in the D-Wave hardware

(e.g. the newer processor has a lower hardware operating temperature, less noise, etc.)

## Classical Benchmarking

### *CD-1 with Alternative Optimizer*

The first step in determining the cause of the better performance from QA was to simply test CD-1 again, now with two optimizers --- momentum (the original) and Adam, a favorite in modern ML contexts. This is to check if the advantage from QA was a result of a simple poor choice of algorithm or if it was somehow a statistical fluke (plausible given the large error bars in the original performance results for CD). The results can be seen in Figure 7, where we show the performance (test accuracy) at 100 and 1000 training epochs for CD-1 both with momentum-based optimization and Adam optimizers as a function of the number of pre-training iterations. Comparing to the results in Figure 1 from the original paper, we see overall performance is broadly similar, and there isn't a significant change in the performance for either optimizer, indicating that it likely wasn't a mere fluke.

### *Contrastive divergence with increased number of steps (CD-k).*

Since we have ruled out chance or optimizers as the origin for the observed advantage, we then move on to the second row of our table and the next phase of the research program – testing classical Gibbs samplers. CD can take a very long time to converge to the true Gibbs distribution of an RBM model, and CD-1 is a very poor approximation in general. D-Wave, on the other hand, has been shown to yield (very) approximate Boltzmann distributions at problem-specific effective temperatures. As such, D-Wave may simply be a better Gibbs sampler. If this is the cause of the performance gain, it would be discovered by exploring better Gibbs samplers on the logical RBM model.

The first step in the investigation of Gibbs sampling as the origin of "quantum" advantage was to test running CD for more Gibbs sampling steps. In Figure 8 we show the performance of CD as implemented in [3] for various numbers of Gibbs steps, from CD-1 to CD-1000. While there may be a slight improvement in the mean accuracy with increasing k, the large error bars make the difference statistically insignificant. Even at 50 pre-training iterations and 1000 post-training iterations, the performance of CD-1000 does not approach the performance of the original QA results.

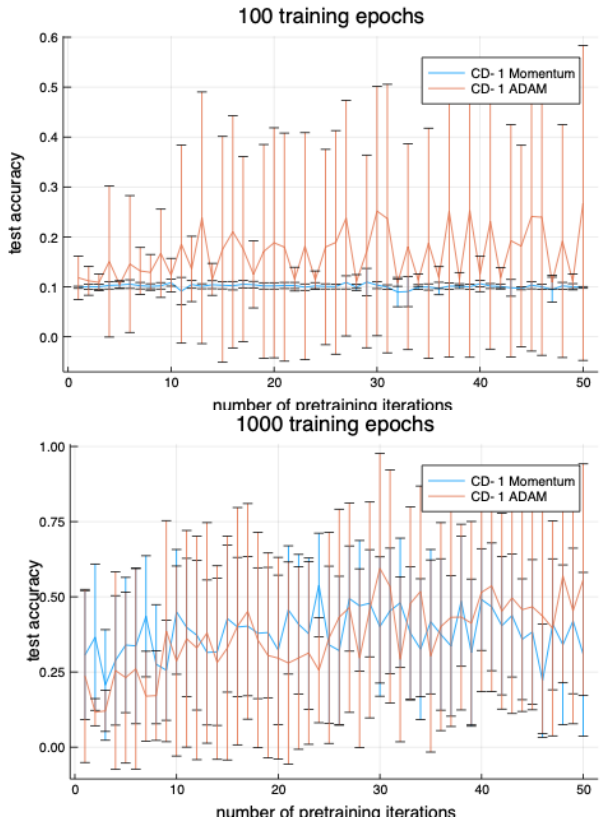

FIG. 7. Performance of CD-1 using momentum (blue) vs. Adam (red) optimizers. Choice of optimizer is evidently irrelevant.

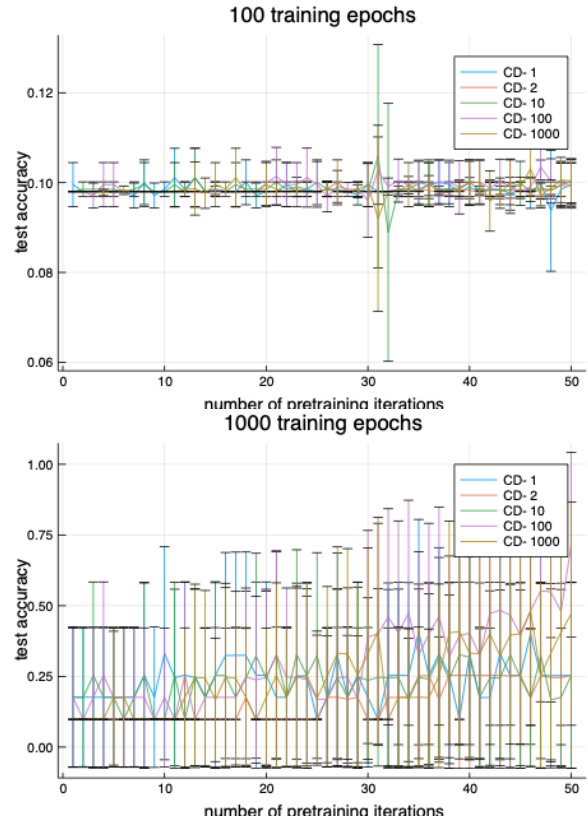

FIG. 8. Performance of CD-k for various k. We see negligible performance improvements through $k = 1000$

*Simulated Annealing*

The next step was to move on to simulated annealing as the simplest thermal sampling beyond straight Gibbs sampling at constant temperature. By slowly adjusting the temperature from a very hot, rapidly thermalizing system to the target temperature, the Monte Carlo Markov chain can converge faster, particularly with multiple restarts (ie multiple runs of the algorithm, which we need anyway to gather statistics to compute the expectation values in the negative phase, such as $\langle v_i h_j \rangle_{\text{model}}$).

We were initially running SA with a large number of sweeps ($10^4$) to more or less guarantee thermalization. Consistent with the hypothesis that the advantage was due to superior thermalization on the quantum annealer, we then experimented with reducing the number of sweeps to see if, as that hypothesis would predict, there was a point where thermalization would fail and performance would degrade significantly, ultimately approximating what was observed with CD-1. Surprisingly, we were able to reduce the number of sweeps all the way down to merely two, as shown in Figure 9.

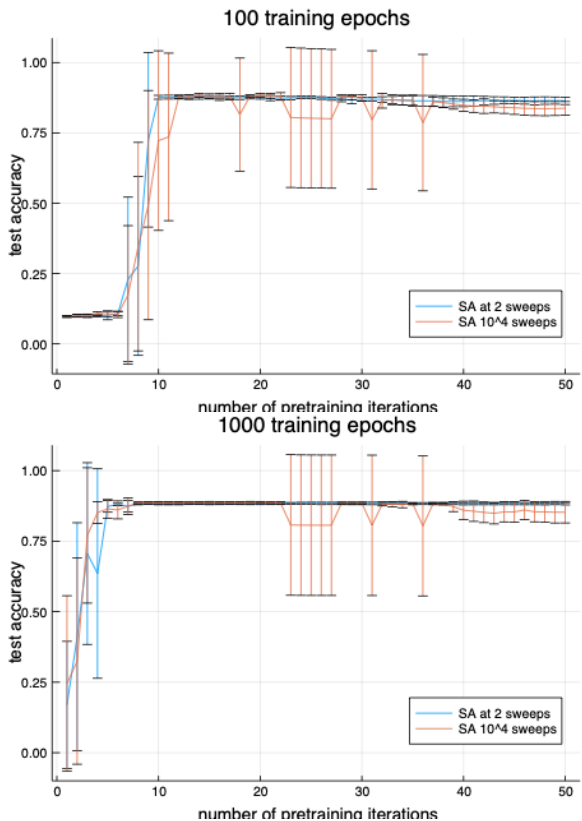

FIG. 9. Performance of networks pretrained with SA at 2 and $10^4$ sweeps, showing SA performs as well as it does asymptotically with merely a quench annealing schedule

Looking at the SA results in Figure 1, we find that there is no statistically significant difference in the distributions of test accuracy as a function of the number of either pre- or post-training iterations between simulated annealing at $10^4$ sweeps versus merely 2. Additional numbers of sweeps were tested but these represent the extremes and all other values behave essentially identically. For both 2 and $10^4$ sweeps, performance quickly rises to ~90%, matching the performance observed for D-Wave in the original paper. Two sweeps is simply one pass at the initial $\beta_{\text{initial}}$ and an immediate quench and single pass at $\beta_{\text{final}}$. This would normally be a very poor approach for thermalization in SA, but in this case it seems to be sufficient.

Thus, it appears that the advantage of quantum-assisted training is not due to inherently quantum effects, but can be reproduced classically using a Gibbs sampler such as SA (even if used in a manner closer to quenching than annealing).

*Contrastive Divergence Revisited.*

It remains somewhat of a paradox however, that SA with just two sweeps is able to reproduce QA performance while CD is not able to do so, even at CD-1000. We considered two possible explanations for this:

1. CD uses block updates (first updating all the hidden nodes, then all the visible nodes, etc) while the SA implementation chooses the next node to be updated randomly.

2. Our SA experiments used 400 independent anneals to estimate the expectation values (this was intended to approximate the D-Wave runs in the original study), while our CD experiments used a number of samples equal to the batch size (100).

To test this, we ran SA with only 100 samples taken per estimate of the negative phase, and also another run, also with 100 samples per estimate, where random variable selection was turned off. In this case, the visible and hidden units were updated in sequence, much like as in CD. The results are in Figure 10 along with those from Figure 9, and as is apparent, both options performed virtually identically to the original/standard SA implementation. Thus, we can categorically rule the above two explanations for the discrepancy between SA and CD.

Fundamentally, this behavior is quite odd --- CD with 1000 steps ought to be able to explore the state space much better than SA with merely 2 sweeps, at least intuitively. Indeed, SA with merely two sweeps is essentially CD-1 where the initial state is random. Thus, we revisited the CD implementation, and rather than using the exact expectation values (ie floating point numbers) for CD, as was done previously, we instead sampled all states, from the initial visible input through to the final configurations of (v,h), from their respective probability densities (given by the activation/sigmoid function). In essence we initialized a purely discrete Markov chain using a value for v sampled from the dataset vector for v (ie $P(v_i = 1) = v_i$ and similarly for $h_j$). The only difference then is initialization --- from a data point vs from a high-temperature approximation of the density, and update method – Metropolis or Gibbs. The results, comparing SA at 2 sweeps against a fully discrete CD implementation at 1 and 1000 Gibbs steps is presented in Figure 11.

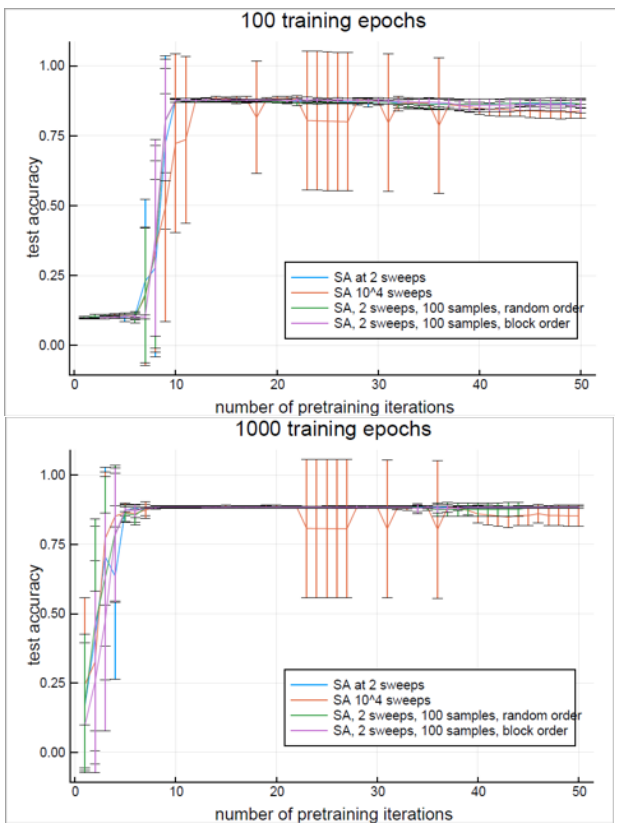

FIG. 10. Results Using SA with varying # of Samples, also block vs. random update order. None of these parameter changes seem to affect the result

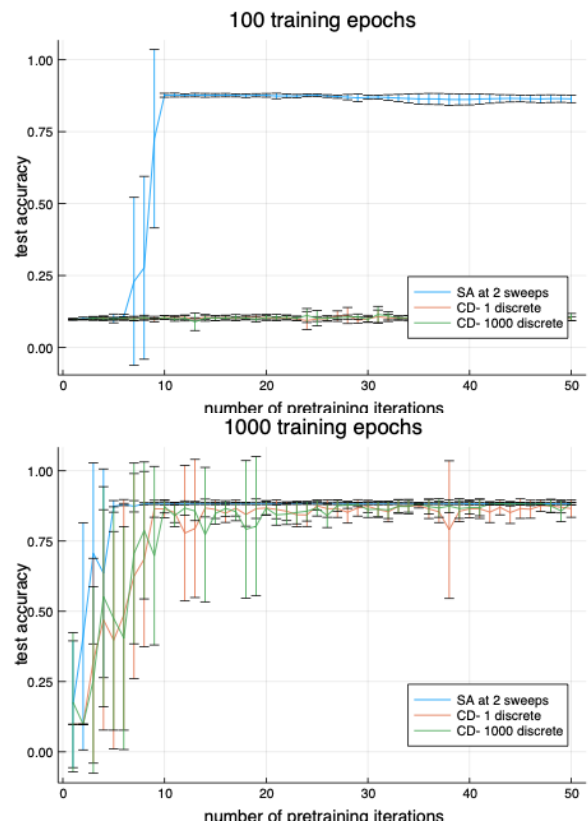

FIG. 11. Performance of networks pretrained with SA (2 Sweeps) vs fully discretized CD, we see a significant convergence advantage for quenched-SA over discretized CD, though we also find CD is able to approach SA performance after a long training period.

As seen in Figure 11, while this completely discretized version of CD is now able to learn about as well as SA after a large number of classical backpropagation training iterations, it does not exhibit the rapid convergence seen in SA or seen previously with QA. As a result, we are forced to conclude that it is the initialization from a random initial state and thermalization at a high-temperature on the energy surface, or the choice of sampling update in the Markov chain (SA uses Metropolis updates, CD Gibbs updates) that yields the advantage. As for the physical reason why any of these would be so important, this remains a mystery, particularly when we compare SA at two sweeps against CD-1000. To truly answer this question would likely require a detailed study of the energy landscape and the dynamics of thermalization, which may be part of a future follow-up study.

## CONCLUSIONS AND LESSONS LEARNED

We reproduced the earlier results for quantum-assisted training on the coarse-grained MNIST data set, which were obtained on a D-Wave 2 processor, using the newer DW2X processor. We then systematically compared the performance of quantum-assisted training vs. a broader range of classical techniques, in order to better understand the underlying reasons for the apparent advantage of the quantum-assisted training over 1-step Contrastive Divergence (CD-1). We found that varying the choice of optimizer or increasing the number of Contrastive Divergence steps as high as CD-1000, did not significantly improve the performance of the classical training. However, we found that using Simulated Annealing (SA) to estimate the model expectations, leads to similar results to quantum-assisted training. Moreover, we were able to achieve these results with SA even using as few as two sweeps, which could better be described as simulated quenching. Thus, it has to be concluded that the apparent advantage of quantum-assisted training is not due to a truly quantum effect, but can be reproduced classically, and in fact by a surprisingly efficient classical algorithm.

We still do not fully understand the reason for the differences in behavior between CD and SA. While CD can be used so as to produce the same ultimate performance after long-time classical backpropagation training, it cannot reproduce the early-time advantage we observed with QA and now again with SA, which is significant --- indeed we find convergence to within 2% of the asymptotic performance of ~89% accuracy after just a dozen pretraining iterations and forty backprop training epochs. After a number of experiments to zero in on the cause, we find that it is localized to the initialization of CD around a datapoint as opposed to the high-temperature initialization from a random point of SA. The physical mechanism for why this yields an advantage is still unclear, but may be of interest for a follow-up study.

At the time the original paper [3] was written, it was recognized that more thorough study was critical to determine whether the apparent advantage of quantum-assisted training on this data set was truly due to quantum effects. Having finally gotten the opportunity to carry out that more systematic investigation, we felt it was important to communicate these results, especially since the original paper has been widely cited and our new study underlines the importance of including a variety of important classical algorithms in one's analyses of quantum annealer performance, not only for optimization, but machine-learning as well. The current study is based on best practices for benchmarking quantum annealers as described in [21]. As we enter the Noisy Intermediate Scale Quantum (NISQ) era and new generations of devices are coming online which are being examined empirically, we hope that similar practices will be followed for benchmarking those devices so that research efforts can be better focused on areas of real quantum advantage.

**Acknowledgments.** This work was supported by the Air Force Research Laboratory under contract FA8750-18-C-0164. We thank P. Alsing, K. Mezzano, and L. Wessing for useful discussions.